\documentclass[runningheads,envcountsame]{llncs}
\usepackage[T1]{fontenc}
\usepackage[utf8]{inputenc}
\usepackage{graphicx}
\usepackage{lineno}
\usepackage{algorithm}

\usepackage{amssymb}
\usepackage{mathtools}

\let\doendproof\endproof
\renewcommand\endproof{~\hfill$\qed$\doendproof}

\usepackage{shortcuts}
\usepackage{listings}

\newif\ifpaper
\paperfalse 

\usepackage{todonotes}

\begin{document}
	\title{Rollercoasters with Plateaus}
	%
	%
	\author{Duncan Adamson\inst{1}\orcidID{0000-0003-3343-2435} \and
		Pamela Fleischmann\inst{2}\orcidID{0000-0002-1531-7970} \and
		Annika Huch\inst{2}\orcidID{0009-0005-1145-5806}}
	\authorrunning{D. Adamson et al.}
	%
	\institute{University of St Andrews \email{duncan.adamson@st-andrews.ac.uk}\and
		Kiel University, Kiel, Germany
		\email{\{fpa,ahu\}@informatik.uni-kiel.de}}
	\maketitle              
    	\begin{abstract}
		In this paper we investigate the problem of detecting, counting, and enumerating (generating) all maximum length plateau-$k$-roller\-coasters appearing as a subsequence of some given word (sequence, string), while allowing for plateaus. We define a plateau-$k$-rollercoaster as a word consisting of an alternating sequence of (weakly) increasing and decreasing \emph{runs}, with each run containing at least $k$ \emph{distinct} elements, allowing the run to contain multiple copies of the same symbol consecutively. This differs from previous work, where runs within rollercoasters have been defined only as sequences of distinct values. Here, we are concerned with rollercoasters of \emph{maximum} length embedded in a given word $w$, that is, the longest rollercoasters that are a subsequence of $w$. 
    	We present algorithms allowing us to determine the longest plateau-$k$-roller\-coasters appearing as a subsequence in any given word $w$ of length $n$ over an alphabet of size $\sigma$ in $O(n \sigma k)$ time, to count the number of plateau-$k$-roller\-coasters in $w$ of maximum length in $O(n \sigma k)$ time, and to output all of them with $O(n)$ delay after $O(n \sigma k)$ preprocessing. Furthermore, we present an algorithm to determine the longest common plateau-$k$-rollercoaster within a set of words in $O(N k \sigma)$ where $N$ is the product of all word lengths within the set.
	\keywords{$k$-Rollercoaster \and Plateaus \and Enumeration \and (Longest Common) Subsequences \and Scattered Factors}
	\end{abstract}

	\section{Introduction}
	Subsequences, also known as scattered factors, of words are heavily studied with various motivations during the last decades:
longest increasing or decreasing (contiguous) subsequence \cite{aldous1999longest,romik2015surprising,schensted1961longest}, 
longest common subsequence \cite{DBLP:journals/jda/KutzBKK11,DBLP:journals/ipl/YangHC05,DBLP:conf/spire/Kosowski04} and 
shortest common supersequence \cite{DBLP:journals/njc/FraserI95,DBLP:journals/siamcomp/JiangL95,DBLP:journals/tcs/RaihaU81}, 
string-to-string-correction \cite{wagner1974string}, and questions related to bioinformatics (e.g. \cite{DBLP:conf/soda/SunW07} 
and the references therein). 
The first two problems can be combined into the notion of {\em rollercoasters} which are  - roughly spoken - sequences such that increasing and decreasing contiguous subsequences alternate and one is interested in the longest rollercoaster as a (contiguous) subsequence of a given sequence (cf. Biedl et al. \cite{DBLP:conf/icalp/BiedlBCLMNS18,DBLP:journals/siamdm/BiedlBCLMNS19}). 
More formally, a \emph{run} is a maximal contiguous subsequence of a given sequence that is strictly increasing or strictly decreasing and, as introduced in \cite{DBLP:conf/gd/BiedlCDJL17}, a \emph{$k$-rollercoaster} is defined as a sequence over the real numbers such that every run is of length at least $k\geq 3$. For instance, the sequence $r=(4,5,7,6,3,2,7,8)$ is a rollercoaster since the runs $(4,5,7)$, $(7,6,3,2)$, and $(2,7,8)$ are increasing resp. decreasing runs of length at least $3$. Regarding the problem of finding the longest rollercoaster in a given sequence, the longest rollercoaster of $(5,3,6,7,1,6,4,1)$
is $(5,6,7,6,4,1)$. Notice that not all sequences contain a rollercoaster witnessed by $(5,3,5,3)$ - all runs are only of length $2$.\looseness=-1

%
Rollercoasters over the real numbers as alphabet are heavily studied. Biedl et al. \cite{DBLP:conf/icalp/BiedlBCLMNS18,DBLP:journals/siamdm/BiedlBCLMNS19} introduced and solved the rollercoaster problem parametrised in the number $\ell\in\N$ of different letters in the given sequence by presenting an $O(\ell \log(\ell))$-time algorithm. Furthermore, they showed that a word of length $n$ does contain a rollercoaster of length at least $\lceil \frac{n}{2} \rceil$ for $n > 7$. In their constructive proof they also gave an algorithm that computes a rollercoaster of this length in $O(n \log( n))$. For a word that is a permutation of $\{1,\ldots, n\}$ they gave an $O(n \log (\log( n)))$ solution.   Gawrychowski et al. \cite{DBLP:journals/algorithmica/GawrychowskiMS22} later improved the runtime of the algorithm such that $k$-rollercoasters can be found in time $O(\ell k^2)$ for sequences with $\ell$ distinct letters. This result shows that longest rollercoasters are related to longest increasing subsequences, that are extensively studied, e.g., \cite{aldous1999longest,DBLP:journals/dm/Fredman75,romik2015surprising,schensted1961longest}.
Moreover, in \cite{DBLP:conf/spire/FujitaNIBT21}, Fujita et al.  solved the \emph{longest common rollercoaster problem} that is to find the longest common rollercoaster contained as a subsequence in two given words by providing two algorithms. The first runs in $O(nmk)$ time and space, where $n, m$ are respectively the lengths of the two words. The second runs in $O(rk\log^3 m \log( \log( m)))$ time and $O(rk)$ space where $r = O(mn)$ is the number of pairs $(i,j)$ of matching positions in the two words that need to contain the same set of letters.\looseness=-1

\noindent
\textbf{Our Contribution.}
Generalising these ideas, we focus our work on rollercoasters with \emph{plateaus}. We refer to a plateau in a rollercoaster as a factor of the rollercoaster containing only a single type of letter. For example, the rollercoaster $(1, 1, 2, 2, 3)$ contains the plateaus $(1, 1)$ and $(2, 2)$, while still being a $3$-rollercoaster. We restrict our definition of $k$-rollercoaster to require that each maximal increasing and decreasing run contains $k$ \emph{unique} letters, rather than simply having length $k$. Therefore, $(1, 1, 2, 2, 3)$ is a plateau-$3$-rollercoaster, however despite containing a weakly increasing subsequence of length $5$, it is not a $5$-rollercoaster.
We focus on three algorithmic questions regarding plateau-$k$-rollercoaster that are related to reachability: first, we give an algorithm that determines the length of the longest plateau-$k$-rollercoaster appearing as a subsequence of a given word of length $n$, and therefore the minimum number of deletions to reach a maximum length plateau-$k$-rollercoaster from a given word of length $n$ in $O(n\sigma k)$ time.
We additionally provide an algorithm for outputting all maximum length plateau-$k$-rollercoasters that can be reached from an input word with at most linear delay relative to the length of the input word. Finally, we give an algorithm to compute the longest common plateau-$k$-rollercoaster within a set given words in $O(N k \sigma)$ time where $N$ is the product of all word lengths within the set.\looseness=-1

%
\noindent
\textbf{Structure of the Work.}
In Section~\ref{prelims}, the basic defintions are presented. In Section~\ref{sec:results} we present the algorithm for computing the longest plateau-rollercoaster within a word and one for enumerating all longest plateau-rollercoasters. Section~\ref{lcr} considers the problem of finding the longest common plateau-rollercoasters within a a set of words.
Due to clear arrangement pseudo code and an extended example can be found in the appendix.\looseness=-1

	\section{Preliminaries}\label{prelims}
	Let $\mathbb{N}$ be the set of all natural numbers, $\mathbb{N}_0 = \mathbb{N} \cup \{0\}$, $[n] = \{1,\ldots,n\}$, 
$[m, n] = \{m, m + 1, \dots, n\}$, and $[n]_0 := [n] \cup \{0\}$ for $m,n \in \N$ with $m \leq n$. 
An \emph{alphabet} $\Sigma$ is a non-empty finite set whose elements are called \emph{letters}. 
We assume w.l.o.g. $\Sigma =\{ 1, 2, \dots, \sigma\}$ for some $\sigma\in\N$, with the usual order on $\N$. 
A \emph{word} over an alphabet $\Sigma$ is a finite sequence of letters from $\Sigma$, with the length of a word $w$, denoted $\vert w \vert$, being the number of letters in $w$.
We define the \emph{empty word} $\varepsilon$ as the word containing no symbols, i.e., $|\varepsilon| = 0$. 
Let $\Sigma^*$ denote the set of all words over $\Sigma$ and $\Sigma^n$ the set of all words of length $n \in \N$. 
Given a word $w \in \Sigma^n$, and integer $i \in [n]$, we denote by $w[i]$ the $i^{th}$ symbol in $w$, and therefore $w = w[1] w[2] \cdots w[n]$. Similarly, given a pair of integers $i, j \in [n]$ such that $i \leq j$, we denote by $w[i, j]$ the word $w[i] w[i + 1] \cdots w[j]$ and set $w[j,i] = \varepsilon$ for $j>i$.
We denote by $\letters(w)$ the set of unique letters in $w$, giving $\letters(w) = \{w[i] \mid i \in [\vert w \vert]\}$
The word $u \in \Sigma^*$ is called a \emph{factor} of $w$ if there exist $x,y \in \Sigma^*$ such that $w = xuy$.
In the case $x=\epsilon$, we call $u$ a \emph{prefix} of $w$ and \emph{suffix} if $y = \epsilon$.
A word $u$ is a \emph{subsequence} (also known as a scattered factor) of $w$ if there exists some set of indices $i_1, i_2, \dots, i_{\vert u \vert} \in [n]$ such that $i_1 < i_2 < \dots < i_{\vert u \vert}$ and $u = w[i_1] w[i_2] \dots w[i_{\vert u \vert}]$. By $w^R$ we denote the reversed word, i.e., $w^R=w[|w|]w[|w|-1]\cdots w[1]$.\looseness=-1

Informally, a plateau-run is a (weakly) increasing or decreasing word. For instance, $w=123345556$ is a plateau-$6$-run since firstly the letters in $w$ are increasing. The subsequence $123456$
is the longest strictly increasing run in $w$. Moreover, $w$ contains the $5$-plateau runs $12334555$ and $23345556$ as well as several shorter plateau-runs. Notice that none of these plateau-runs is maximal. The word $544133465$ contains the maximal decreasing $3$-plateau-run $5441$ and the maximal increasing $4$-plateau-run $13346$.\looseness=-1

\begin{definition}\label{prun}
	A factor $u = w[i, j]$ of $w \in \Sigma^n$ is a \emph{plateau-run} if, for all $\ell \in [\vert u \vert -1]$, we either have $u[\ell] \leq u[\ell +1]$ or $u[\ell] \geq u[\ell+1]$. The \emph{orientation} of a run is $\uparrow$ (\emph{increasing}) if $u[\ell] \leq u[\ell + 1]$, or $\downarrow$ (\emph{decreasing}) if $u[\ell] \geq u[\ell + 1]$ for every $\ell \in [\vert u \vert - 1]$.
	Such a plateau-run is called {\em maximal} if $w[i - 1] > u[1]$ and $u[\vert u \vert] > w[j + 1]$ or 
 $w[i - 1] < u[1]$ and $u[\vert u \vert] <w[j + 1]$ resp. (notice that if $u$ is a prefix or suffix of $w$ the related constraints on the maximality are omitted).\looseness=-1
 %
	A plateau-run $u$ is a plateau-$k$-run for some $k\in\N$ if we have $\vert\letters(u)\vert \geq k$.\looseness=-1
\end{definition}

Given a variable $\xi \in \{\uparrow, \downarrow\}$, we use $\overline{\xi}$ to denote the opposite orientation, i.e., $\overline{\uparrow} = \downarrow$ and $\overline{\downarrow} = \uparrow$.\looseness=-1

\begin{remark}
Notice that one obtains the classical {\em run} introduced in  
\cite{DBLP:conf/icalp/BiedlBCLMNS18,DBLP:journals/siamdm/BiedlBCLMNS19} by changing $\leq$ and $\geq$ in Definition \ref{prun} into $<$ and $>$.\looseness=-1
\end{remark}

We define rollercoasters in our setting as a class of words containing alternating (weakly) increasing and decreasing factors that are neither left- nor right-extendable and all contain at least 3 distinct letters.
For instance, $12345435667$ is a plateau-$3$-rollercoaster consisting of the maximal plateau-$3$-runs $12345$, $543$, $35667$ while $1234554567$ is not a plateau-$3$-rollercoaster since $554$ is a maximal run but not a plateau-$3$-run.\looseness=-1

\begin{definition}
A word $w \in \Sigma^*$ is a \emph{plateau-$k$-rollercoaster} for $k \in \mathbb{N}, k \geq 3$ if every plateau-run $r$ in $w$ is a maximal plateau-$k$-run.\looseness=-1
\end{definition}

\begin{remark}
	Note that the required maximality of the runs within a plateau-rollercoaster implies that their orientation is alternating, i.e., if the \nth{$i$} run of $w$ is an increasing run, then $w$'s \nth{$(i+1)$} run is decreasing and vice versa.\looseness=-1
\end{remark}

\begin{remark}
First, notice that plateau-$k$-rollercoasters are an extension of classical rollercoasters by allowing plateau-$k$-runs instead of classical runs. In contrast to the classical rollercoaster, in plateau-$k$-rollercoasters the runs may overlap in more than one letter. Consider, for instance, the plateau-$4$-rollercoaster
$w=12223444321112345$ with the plateau-runs $r_1=12223444$, $r_2=44432111$, and $r_3=1112345$. In order to
decompose a rollercoaster into its runs we define the concatenation $x_1\pdot x_2$ by $x_1p^{-1}x_2$ where $p$ is the longest prefix of $x_2$ which is also a suffix of $x_1$, for $x_1,x_2\in\Sigma^{\ast}$, and $p^{-1}x_2=y$ iff $x_2=py$.\looseness=-1
\end{remark}

We extend the notion of $(k,h)_w$-rollercoasters as introduced by Biedl et al. 
\cite{DBLP:conf/icalp/BiedlBCLMNS18,DBLP:journals/siamdm/BiedlBCLMNS19} by not counting the length of the last run but the number of letters that are part of the strictly increasing or decreasing subsequence of the last run. This notion will be mainly used within the algorithmic constructions of rollercoasters. Here we follow the idea of forming a rollercoaster by successively appending letters and tracking whether we completed a $k$-run (which is done by the variable $h$).\looseness=-1

\begin{definition}
	For $\xi \in \{\uparrow,\downarrow\}$, $k \in \N$ and $h,\ell \in [k]$, we say that a word $w$ is a plateau-$(k,h)_\xi$-rollercoaster if the following properties hold for the decomposition $w=r_1\pdot \ldots\pdot r_x$, $x\in \N$ of $w$ into runs:\looseness=-1
	\begin{enumerate}
		\item The last plateau-run $r_x$ has orientation $\xi$.\looseness=-1
		\item For all $i\in [x-1]$ the $r_i$ is a plateau-$k$-run. The last plateau-run $r_x$ is a plateau-$h$-run if $h\in[k-1]$ and a plateau-$k$-run if $h=k$.\looseness=-1
	\end{enumerate}
\end{definition}

\begin{remark}\label{rem:kkandk1rollercoasters}
Note that a plateau-$(k,k)_\xi$-rollercoaster for $\xi \in \{\uparrow,\downarrow\}$ is a classic plateau-$k$-rollercoaster. Further, note that a plateau-$(k,1)_\uparrow$-rollercoaster is either a unary word or also a plateau-$(k,k)_\downarrow$-rollercoaster since neighboured runs do overlap in in their respective ends/beginnings within unary factors (analogously for plateau-$(k,1)_\downarrow$-rollercoaster).\looseness=-1
\end{remark}

The word $r = 12234322$ is not a plateau-$4$-rollercoaster since its last run does only contain three distinct letters. Since the last run has orientation $\downarrow$, $r$ is a plateau-$(4,3)_\uparrow$-rollercoaster.
Further, consider $w = 43321$ and $k = 3$ which is a plateau-$(3,3)_\downarrow$-rollercoaster but also a plateau-$(3,1)_\uparrow$-rollercoaster since $w$'s last letter not only belongs to the decreasing run but also starts a new increasing run of of length 1 itself.\looseness=-1

For the remainder of this work, we are interested in plateau-$k$-rollercoasters that appear as subsequences of some given word $w$, utilising plateau-$(k,h)_\xi$-rollercoasters as a major tool in our algorithms. 
For these algorithmic results we use the standard 
computational model RAM with logarithmic word-size (see, e.g., \cite{KarkkainenSB06}), i.e., we follow a standard assumption from stringology, that if $w$ is the 
input word for our algorithms with $\Sigma=\letters(w)=\{ 1,2, \ldots, \sigma\}$.\looseness=-1

	\section{Counting and Enumerating Plateau-$k$-Rollercoasters}\label{sec:results}
	In this section we present our results regarding detecting, counting, and enumerating the set of plateau-$k$-rollercoasters that appear as a subsequence within a given word $w\in\Sigma^{n}$ for some $n \in \mathbb{N}$. Thus, we are extending the classical rollercoaster problem - given a word $w\in\Sigma^{\ast}$, determine the longest subsequence of $w$ which is a rollercoaster - to the plateau-rollercoaster scenario. For better readability, we say that a rollercoaster $r$ is in a word $w$, if $r$ is indeed a rollercoaster and additionally a subsequence of $w$.\looseness=-1

We start with an algorithm for determining the longest plateau-$k$-rollercoaster in a given word $w \in \Sigma^n$. To do so, we introduce two set of tables. First, the set of tables $L_{w}^{k, h,\xi}$ (\textbf{L}ongest plateau-$(k,h)_{\xi}$-rollercoaster), with $L_{w}^{k, h,\xi}[i]$ denoting the length of the longest plateau-$(k,h)_{\xi}$-rollercoaster in $w[1, i]$ ending at position $i$, noting that this may be different from the length of the longest plateau-$(k,h)_{\xi}$-rollercoaster in $w[1, i]$. We abuse our notation by using $L_{w}^{k, k,\xi}[i]$ to denote the length of any proper plateau-$k$-rollercoaster ending with a $\xi$-run at $w[i]$. Secondly, we have the $n \times \sigma$ table $P$ where $P_w[i,x]$ contains the index $i'$ such that $i' \in [1, i]$ where $w[i'] = x$ and, $\forall j \in [i' + 1, i]$, $w[j] \neq x$.\looseness=-1

\begin{remark}
Notice that $P_w$ can be constructed in time $O(n\sigma)$ for a given $w\in\Sigma^n$.\looseness=-1
\end{remark}

For $w = 871264435161$, we have  $L_w^{3,3,\uparrow}[9] = 7$ since the longest plateau-$(3,3)_\uparrow$-rollercoaster ending in $w[9]$ is $8712445$. 
As a second example with an incomplete run in the end, we have $L_w^{3,2,\downarrow}(w)[8] =7$ witnessed by the plateau-$(3,2)_\downarrow$-rollercoaster $8712443$ ending in $w[8]$. 
The full exemplary tables for $P_w$ and $L_w^{3,h,\xi}$ for $h \in [3]$, $\xi \in \{\downarrow,\uparrow\}$ can be found in Appendix~\ref{app:example}. There we can see that the longest plateau-$3$-rollercoasters that can be reached via deletions from $w$ is of length $10$ are given by $8712644311$ and $8712644356$. Furthermore, the relation between $L_w^{k,1,\xi}$ and $L_w^{k,k,\overline{\xi}}$ (cf. Remark~\ref{rem:kkandk1rollercoasters}) gets perfectly visible since the respective rows do either represent unary plateau-rollercoasters or contain equal values.
Using dynamic programming, we can compute the length of the longest common plateau-$k$-rollercoaster that ends in position $i$ in a word $w$.\looseness=-1

\begin{lemma}
    \label{lem:L_k_h_table}
    Given a word $w \in \Sigma^n$ and $i, k, h \in [n]$, then $L_w^{k, h,\xi}[i]$ can be determined in $O(\sigma)$ time from  $L_w^{k, h',\xi'}[j]$, for all $\xi' \in \{\uparrow, \downarrow\}, h' \in [k], j \in [i - 1]$.\looseness=-1
\end{lemma}

\ifpaper
\else
\begin{proof}
    Observe that the longest plateau-$(k,h)_{\xi}$-rollercoaster ending at $w[i]$ must contain, as a prefix, 
    some plateau-$(k,h')_{\xi'}$-rollercoaster ending at $w[j]$ for some $j \in [1, i -1 ]$, $h' \in [1, k]$ and $\xi' \in \{\uparrow, \downarrow\}$. Further, $j = P_w[i - 1, w[j]]$, as otherwise there existed a longer plateau-$(k,h')_{\xi'}$-rollercoaster ending at $P_w[i - 1, w[j]]$.  
    Let us assume that $\xi = \uparrow$ and note that the arguments are analogous for $\xi = \downarrow$. 
    
    If $h \in [2, k - 1]$, then we are looking for the value $j$ such that 
    \begin{itemize}
    \item $w[j] \leq w[i]$ and 
    \item for all $j' \in [1, i - 1]$ with $w[j'] \leq w[i]$, 
        \begin{itemize}
        \item either $L_w^{k, h - 1,\uparrow}[j'] < L_w^{k, h - 1,\uparrow}[j]$ holds, if $w[j] \neq w[i]$, 
        \item or $L_w^{k, h - 1,\uparrow}[j'] < \max\{L_w^{k, h,\uparrow}[j], L_w^{k, h - 1,\uparrow}[j]\}$ holds if $w[j] = w[i]$. 
        \end{itemize}
    \end{itemize}
    Note that this can be found by checking, in $O(\sigma)$ time, each value of $j \in \{P_w[i - 1, x] \mid x \in \Sigma\}$ sequentially and choosing the value maximising either $L_w^{k, h - 1,\uparrow}[j]$, if $w[j] < w[i]$ or $\max\{L_w^{k, h,\uparrow}[j], L_w^{k, h - 1,\uparrow}[j]\}$ if $w[j] = w[i]$. 
    
    Otherwise, if $h = 1$, then as $w[i]$ is the first element of a $\xi$-run, it must also be the last element of a $\overline{\xi}$-run, or continuing a plateau of the last element of such a run. As such, we are instead looking for the index $j \in \{P_w[1, x] \mid x \in \Sigma\}$ maximising 
    \begin{itemize}
    \item     either $\max\{L_w^{k, k,\downarrow}[j], L_w^{k, k - 1,\downarrow}[j]\}$, if $w[j] > w[i]$, 
    \item or $\max\{L_w^{k, k,\downarrow}[j], L_w^{k, k - 1,\downarrow}[j], L_{\uparrow}^{k, 1}(w)[j]\}$ if $w[j] = w[i]$. 
    \end{itemize}
    Again, this can be determined in $O(\sigma)$ time by checking each value of $j \in \{P_w[i - 1, x] \mid x \in \Sigma\}$ sequentially. 
    
    Finally, if $h = k$, then $w[i]$ is either the \nth{$k$} symbol of a run containing $k - 1$ letters, the continuation of a plateau of a run containing at least $k$ unique letters, or an extension of a run containing at least $k$ unique letters. Therefore, the problem becomes determining the value of $j \in \{P_w[i - 1, x] \mid x \in \Sigma\}$ maximising 
    \begin{itemize}
    \item $\max\{L_w^{k, k - 1,\uparrow}[j], L_w^{k, k,\uparrow}[j]\}$, if $w[j] < w[i]$, 
    \item or $L_w^{k, k,\uparrow}[j]$ if $w[j] = w[i]$, 
    \end{itemize}
    the value of which may be determined in $O(\sigma)$ time in the same manner as before.
\end{proof}

\fi


\begin{corollary}
    \label{col:full_L_j_h_table}
    Given a word $w \in \Sigma^n$, the value of $L_w^{k, h,\xi}[i]$ can be determined in $O(n k\sigma)$ time for every $\xi \in \{\uparrow, \downarrow\}, h \in [1, k], i \in [1, n]$.\looseness=-1
\end{corollary}

\begin{theorem}
    \label{thm:length_of_the_longest_roller_coaster}
    Given a word $w \in \Sigma^n$, we can determine the length of the longest plateau-$k$-rollercoaster in $O(n \sigma k)$ time.\looseness=-1
\end{theorem}
\begin{proof}
    From Corollary \ref{col:full_L_j_h_table}, we can construct the table $L_w^{k, h,\xi}$ in $O(n \sigma k)$ time. By definition, the longest rollercoaster is the value $\max_{i \in [1, n]} L_w^{k, k,\xi}[i]$, which may be determined in $O(n)$ time from $L_w^{k, h,\xi}$, giving the stated time complexity.\looseness=-1
\end{proof}

Now, we consider the problems of counting and enumerating the set of maximum-length plateau-rollercoasters in $w$. To do so, we define for a word $w$ the \emph{$k$-roller\-coaster table}, $R_w$. Informally, $R_w$ can be thought of as an extension of $L_w^{k, h,\xi}$, storing not only the longest plateau-$(k,h)_{\xi}$-rollercoasters ending at each position, but also the number of such rollercoasters.\looseness=-1

\begin{definition}[Rollercoaster Table]
    \label{def:rollercoaster_table}
    Given a word $w\in\Sigma^n$, for some $n \in \N$, and $k\in\N$, the \emph{rollercoaster table} of $w$ and $k$ is the table $R_w^k$ of size $n \times 
    2 \times k$, indexed by the triples $i \in [1, n]$, $\xi \in \{\uparrow,
    \downarrow\}$, and $h \in [1, k]$, where $R_w^k[i, \xi, h]$ is the number of subsequences $s$ which are plateau-$(k,h)_{\xi}$-rollercoasters
    ending at position $i$ in $w$, i.e. $w[i]$ is the last symbol of $s$.\looseness=-1
\end{definition}

We now outline our approach for computing $R_w^k$, providing pseudocode for this algorithm in Algorithm \ref{alg:computing_R_w} in Appendix \ref{app:pseudocode} and a proof of correctness in Theorem \ref{thm:computing_R_w}, using the table $L_w^{k, h,\xi}$ as a basis.\looseness=-1

\begin{theorem}
    \label{thm:computing_R_w}
    There exists an algorithm computing $R_w^k$ for a given input word $w \in \Sigma^{n}$ and $k\in\N$  in $O(n \sigma k)$ time.\looseness=-1
\end{theorem}

\ifpaper
\else
\begin{proof}
    We prove this statement using a dynamic programming approach. As a base case, note that $R_w^k[i, \xi, 1]$ is equal to $1$ for every $i \in [n]$, $\xi \in \{\uparrow, \downarrow\}$, corresponding to the plateau-$(k,1)_{\xi}$-rollercoaster associated to position $i$, which may be either an increasing or decreasing run. In general, the value of $R_w^k[i, \xi, h]$ can be computed by one of six summations, depending on the values of $\xi$ and $h$.

    If $h \in [2, k - 1]$, then $R_w^k[i, \uparrow, h]$ is equal to the sum of the number of plateau-$(k, h - 1)_\uparrow$-rollercoasters of length $L_w^{k, h - 1,\uparrow}[i] - 1$ that end on a symbol smaller than $w[i]$, plus the number of plateau-$(k, h)$-rollercoasters of length $L_w^{k, h,\uparrow}[i] - 1$ and ending with the symbol $w[i]$. Define
    the array $\PP_w^{k,h,\xi}$ of length $n$ by
    \[
    \\PP_w^{k,h,\uparrow}[i] = \{P_w[i - 1, x] \mid x \in [1, w[i] - 1], L_w^{k, h - 1,\uparrow}[P_w[i - 1, x]] = L_w^{k, h,\uparrow}[i] - 1\}
    \]
    for all $i\in[n]$. Then, we get
    \begin{multline*}
        R_w^k[i, \uparrow, h] = \left(\sum\limits_{i' \in PP^{k, h,\uparrow}_w[i]} R_w^k[i', \uparrow, h - 1] \right) +\\
       \begin{cases}
            R_w^k[P_w[w[i], i - 1], \uparrow, h] & L_w^{k, h,\uparrow}[i] - 1 = L_w^{k, h,\uparrow}[P_w[w[i], i - 1]]\\
            0 & L_w^{k, h,\uparrow}[i] - 1 \neq L_w^{k, h,\uparrow}[P_w[w[i], i - 1]]
        \end{cases}.
    \end{multline*}
    
    An analogous summation can be derived for computing $ R_w^k[i, \downarrow, h]$ by defining
    \[
    PP^{k, h,\downarrow}_w[i] = \{P_w[i - 1, x] \mid x \in [w[i] + 1, \sigma], L_w^{k, h - 1,\downarrow}[P_w[i - 1, x]] = L_w^{k, h,\downarrow}[i] - 1\}.
    \]

    If $h = k$ then the value of $R_w^k[i, \uparrow, k]$ is equal to the number of plateau-$(k, k - 1)_{\uparrow}$-rollercoasters of length $L_w^{k, k,\uparrow}[i] - 1$ ending before position $i$ with any symbol smaller than $w[i]$, plus the number of plateau-$(k, k)_{\uparrow}$-rollercoasters of length $L_w^{k, k,\uparrow}[i] - 1$ ending before position $i$ with any symbol less than or greater than $w[i]$. Note the the first set corresponds to the rollercoasters in the set $\PP_w^{k, k - 1,\uparrow}[i]$, while the second is equal to the set 
    \[
    \PP^{k, h,\uparrow}_w[i]' = \{P_w[i - 1, x] \mid x \in [1, w[i]], L_w^{k, h - 1,\uparrow}[P_w[i - 1, x]] = L_w^{k, h,\uparrow}[i] - 1\}, 
    \]
    leading to
    \[
        R_w^k[i, \uparrow, k] = \left(\sum\limits_{i' \in PP^{k, k,\uparrow}_w[i]} R_w^k[i', \uparrow, k - 1] \right) + \left(\sum\limits_{i' \in PP^{k, k,\uparrow}_w[i]'} R_w^k[i', \uparrow, k] \right).
    \]

    Again, an analogous summation may be derived for $R_w^k[i, \downarrow, k]$ by replacing the array 
    $\PP_w^{k, k,\uparrow}$ by $\PP_w^{k, k,\downarrow}$ and $(\PP_w^{k, k,\uparrow})'$ by 
    \[
    \PP^{k, k,\downarrow}_w[i]' = \{P_w[i - 1, x] \mid x \in [w[i], \sigma], L_w^{k, h - 1,\downarrow}[P_w[i - 1, x]] = L_w^{k, h,\downarrow}[i] - 1\}.
    \]
    Finally, the value of $R_w^k[i, \xi, 1]$ is exactly equal to the size of $\max\{1, R_w^k[i, \overline{\xi}, k]\}$ by definition.

    To determine the complexity of this computation, note that we can, as a base case, set the value of $R_w^k[1, \xi, 1]$ to $1$, and of $R_w^k[1, \xi, h]$ to $0$, for every $h \in [2, k]$. Now, note that the sets $PP^{k, h,\uparrow}_w[i]$, $PP^{k, h,\downarrow}_w[i]$, $PP^{k, k,\uparrow}_w[i]'$, and $PP^{k, k,\downarrow}_w[i]'$ can each be computed in $O(\sigma)$ time assuming the values of $L_{w}^{k, h',\uparrow}[i']$ have been computed for every $h \in [1, k]$ and $i' \in [1, n]$. Then, using the above summations, we can compute the value of $R_w^k[i, \xi, h]$ in a further $O(\sigma)$ time, assuming that $ R_w^k[i', \xi', h']$ has been computed for every $i' \in [1, i - 1], \xi' \in \{\uparrow, \downarrow\}$, and $h' \in [1, k]$. As there are $n k$ entries in the table $R_w^k$, we have a total time complexity of $O(n k \sigma)$.\looseness=-1
\end{proof}

\fi

Regarding the problems of counting and enumerating plateau-rollercoasters, we need show that the longest common plateau-rollercoaster can only be found at a unique position within $w$.\looseness=-1

\begin{proposition}\label{obs:uniqueness_of_maximum_rollercoasters}
    Given a word $w \in \Sigma^n$ and pair of plateau-rollercoasters $s$ and $s'$ such that $s = w[i_1] w[i_2] \dots w[i_m]$ and $s' = w[j_1] w[j_2] \dots w[j_m]$ where $m$ is the length of the longest plateau-rollercoaster that is a subsequence of $w$, either $s \neq s'$ or $(i_1, i_2, \dots, i_m) = (j_1, j_2, \dots, j_m)$.
\end{proposition}
\begin{proof}
    If $(i_1, i_2, \dots, i_m) = (j_1, j_2, \dots, j_m)$ then clearly $s = s'$. Assuming that $(i_1, i_2, \dots, i_m) \neq (j_1, j_2, \dots, j_m)$, we have that $s = s'$ if and only if $w[i_c] = w[j_c], \forall c \in [1, m]$. Let $t \in [1, m]$ be the value such that $i_1, \dots, i_{t - 1} = j_1, \dots, j_{t - 1}$ and $i_t \neq j_t$. If $s = s'$ then $w[i_t] = w[j_t]$. If $i_t < j_t$ then $w[i_1] \dots w[i_t]  \dots w[j_{m}]$ must be a plateau-rollercoaster of length $m + 1$, contradicting the assumption that $m$ is the length of the longest plateau-rollercoaster in $w$. Analogously, if $i_t > j_t$, $w[j_1] w[j_2] \dots w[j_t] w[i_t] w[i_{t + 1} \dots w[i_{m}]$ must be a plateau-rollercoaster. Hence the claim holds.
\end{proof}

\begin{lemma}
    \label{lem:counting_maximum_rollercoasters}
    Let $w \in \Sigma^n$ for $n\in\N$ be a word where the length of the longest plateau-rollercoaster in $w$ is $m\in\N$. Then, the total number of plateau-rollercoas\-ters in $w$ of length $m$ is given by
    \[
    	\sum\limits_{i \in [1, n], \xi \in \{\uparrow, \downarrow\}} R_w^k[i, \xi, 0].
    \]
    
\end{lemma}
\begin{proof}
    Following Lemma \ref{obs:uniqueness_of_maximum_rollercoasters}, any plateau-rollercoaster of maximum length $m$ in $w$ must be unique. Therefore, given any pair of indices $i, i' \in [1, n]$, any $m$-length plateau-rollercoaster ending at $w[i]$ must be distinct from any such plateau-rollercoaster ending at $w[i']$. Hence the above statement holds.
\end{proof}



In order to enumerate the set of plateau-rollercoasters in a given word $w$, we need one additional auxiliary structure, the \emph{next element graph} of $w$, denoted $\NEG(w)$. Here every node represents a position of $w$, precisely $v_i$ represents $w[i]$ for all $i\in[|w|]$.\looseness=-1

\begin{definition}
For a given word $w\in\Sigma^n$ for $n\in\N$, define $\NEG(w) = (V, E)$ as the edge-labeled, directed graph with 
$V = \{v_1, v_2, \dots, v_n\}$, $E=\{(v_i,v_j)|\,i\in[n-1],j\in[i+1,n]\}$, and the edge labelling function $\ell:E\rightarrow\{\uparrow,\rightarrow,\downarrow\}$ with
\[
\ell(e)=
\begin{cases}
\uparrow,&\mbox{if } w[i] < w[j]\land \forall j' \in [i + 1, j - 1]:\,  w[i] > w[j']\lor w[j'] > w[j],\\
\rightarrow,&\mbox{if } w[i] = w[j] \land i = P_w[j - 1, w[j]],\\
\downarrow,&\mbox{if } w[i] > w[j] \land \forall j' \in [i + 1, j - 1]:\, w[i] < w[j] \lor w[j'] < w[j].
\end{cases}
\]
\end{definition}

\begin{lemma}
    \label{lem:constructing_NEG}
    Given a word $w \in \Sigma^n$, $\NEG(w)$ can be constructed in $O(n \sigma)$ time (cf. Algorithm~\ref{alg:constructing_NEG}).\looseness=-1
\end{lemma}
\ifpaper
\else
\begin{proof}
    We achieve this by working backwards, adding in-edges to each vertex. Notice that our construction does not rely on the ordering of the vertices, and therefore it is sufficient to prove that our construction adds all incoming edges to the current vertex $v$.\\
    \textbf{Claim.} Given the vertex $v_j$, we claim that 
    there exists some edge starting at $v_i$ and ending at $v_j$ labelled by $\uparrow$ if and only if $i \in \{P_w[j - 1, x] \mid x \in [1, w[j] - 1], \forall x' \in [x + 1, w[j]], P_w[i - 1, x'] < P_w[i - 1, x]\}$. \\
    Observe that, by the definition of $\NEG(w)$, any position $i'$ not in this set either contains a symbol greater than or equal to $w[j]$, and thus can not be used as a unique symbol in an increasing run, or is followed by some $x$ such that $w[i'] \leq x \leq w[j]$, and therefore a longer increasing run can be formed by inserting $x$ between $w[i]$ and $w[j]$. In the other direction, given any $i \in \{P_w[j - 1, x] \mid x \in [1, w[j] - 1], \forall x' \in [x + 1, w[j]]: P_w[i - 1, x'] < P_w[i - 1, x]\}$, any $i' \in [i + 1, j - 1]$ satisfies either $w[i'] < w[i]$ or $w[i'] > w[j]$, and hence the edge $(v_i, v_j)$ labelled $\uparrow$ exists in $\NEG(w)$.
    
    \medskip
    
    We construct the set 
    \[
    \{P_w[j - 1, x] \mid x \in [1, w[j] - 1], \forall x' \in [x + 1, w[j]], P_w[i - 1, x'] < P_w[i - 1, x]\}
    \]
    in $O(\sigma)$ time by checking each $x \in [1, w[j] - 1]$ in decreasing order if $P[j - 1, x] < P_w[i - 1, x']$ for all $x' \in [x + 1, w[j]]$. By storing some value $p$ corresponding to $\max_{x' \in [x + 1, w[j]]} P_w[j - 1, x']$, initially set to $P_w[j - 1, w[j]]$, we can determine if this holds in $O(1)$ time, updating $p$ to $\max\{P_w[j - 1, x], p\}$ after checking each value of $x$, allowing the set to be constructed in $O(\sigma)$ time, and thus all in-edges ending at $v[j]$ labelled by $\uparrow$ to be determined in $O(\sigma)$ time. By analogous techniques, we can determine the set of in-edges ending at $v[j]$ labelled by $\downarrow$ in $O(\sigma)$ time. Finally, the edge labelled $\rightarrow$ can be determined from $P_w$ in constant time, and therefore $\NEG(w)$ can be constructed in $O(n \sigma)$ time.
\end{proof}

\fi

\begin{algorithm}
\textbf{Input:} $w \in \Sigma^n$, $k \in \mathbb{N}$, Rollercoaster table $R^k_w$\\
\textbf{Output:} graph $NEG(w)$
\begin{lstlisting}[mathescape=true,basicstyle=\tiny]
Initialise $L_w^{k, h,\xi}$, $\forall \xi \in \{\uparrow, \downarrow\}, h \in [1, k]$
Compute $P_w$
Set $V = \{ v_1, v_2, \dots, v_n\}$
Set $E = \emptyset$ // edges are triples $(v_i, v_j, \xi)$
// iteration over vertices to add the edges ($v_1$ has no incoming edges)
for $i \in [2, n]$ 
  $p_{\uparrow} = P_w[i - 1, w[i]]$ // initial bound on which $\uparrow$-labelled edges are blocked by existing edges, we do not need to explicitly construct the set 
  for $x \in \{w[i] - 1, \dots, 1\}$
    if $P_w[i - 1, x] > p_{\uparrow}$ // if $P_w[i - 1, x]$ after every symbol in $w$ 
       $E = E \cup \{ (v_{P_w[i - 1, x]}, v_i, \uparrow)\}$
       $p_{\uparrow} = \max(p_{\uparrow}, P_w[i - 1, x])$
  $p_{\downarrow} = P_w[i - 1, w[i]]$ // We now do the same for the edges labelled $\downarrow$
  for $x \in [w[i] + 1,  \sigma]$
    if $P_w[i - 1, x] > p_{\downarrow}$ // if $P_w[i - 1, x]$ after every symbol in $w$ 
       $E = E \cup \{ (v_{P_w[i - 1, x]}, v_i, \downarrow)\}$
       $p_{\downarrow} = \max(p_{\downarrow}, P_w[i - 1, x])$
  // add labelled $\rightarrow$ from $v_{P_w[i - 1, w[i]}$ to $v_i$ 
  $E = E \cup \{(v_{P_w[i - 1, w[i]]}, v_i, \rightarrow)\}$  
return $(V, E)$
\end{lstlisting}
\caption{Construct$NEG$}\label{alg:constructing_NEG}
\end{algorithm}

    First, we show that we can associate a path in $\NEG(w)$ to unique subsequences of $w$.\looseness=-1

\begin{lemma}
    \label{obs:every_path_is_unique}
    Let $w\in\Sigma^n$ for some $n\in\N$.
Moreover, let $i_1, i_2, \dots, i_{\sigma}$ be the set of position such that $w[i_q] = q$ and for all $i' \in [1, i_q - 1], w[i'] \neq q$, i.e., $i_q$ is the left most occurrence of the letter $q$ in $w$. Every path of length $j$ in $\NEG(w)$ starting at any vertex $v_{i_q}$ corresponds to a unique subsequence of $w$.\looseness=-1
\end{lemma}
\ifpaper
\else 
\begin{proof}
    Assume, for the sake of contradiction, that there exists some pair of paths $P = (v_{p_1}, v_{p_2}, \dots, v_{p_j})$, and $Q =( v_{q_1}, v_{q_2}, \dots, v_{q_j})$, where $w[p_1] w[p_2] \dots w[p_j] = w[q_1] w[q_2] \dots w[q_j]$. Now, as $w[p_1] = w[q_1]$, both $P$ and $Q$ must start at vertex $v_{w[p_1]}$, so $p_1 = q_1$. Similarly, observe that if the edge $(v_{p_1}, v_{p_2})$ exits in $\NEG(w)$, and $w[p_2] = w[q_2]$, then $(v_{p_1}, v_{q_2})$ exists in $\NEG(w)$ if and only if $p_2 = q_2$. In general, if $p_1, p_2, \dots, p_{i - 1} = q_1, q_2, \dots, p_{i - 1}$, as $w[p_i] = w[q_i]$, we have that if the edge $(v_{p_{i - 1}}, v_{p_i})$ exists in $\NEG(w)$, the edge $(v_{p_{i - 1}}, v_{q_i})$ is in $\NEG(w)$ if and only if $p_i = q_i$, completing the proof.
\end{proof}

\fi

Now, we prove that every maximum length plateau-rollercoaster in $w$ corresponds to a path in $\NEG(w)$.\looseness=-1

\begin{lemma}
    \label{lem:every_maximum_length_rollercoaster_is_a_path}
    Given a word $w \in \Sigma^n$ for $n\in\N$ and a maximal plateau-rollercoas\-ter $u$ in $w$, there exists exactly one path $(i_1, i_2, \dots, i_{\vert u \vert})$ in $\NEG(w)$  such that $u=w[i_1] w[i_2] \dots w[i_{\vert u \vert}]$.\looseness=-1
\end{lemma}
\begin{proof}
    Let $i_1$ be the value such that for all $i \in [1, i_1 - 1]$ we have $w[i] \neq u[1]$ and $u[1] = w[i_1]$, and $i_2$ be the value such that for all $j \in [i_1 + 1, i_2 - 1]$, we have $u[2] \neq w[j]$. As $u$ is maximal in $w$, there must be the edge $(v_{i_1}, v_{i_2})$ in $\NEG(w)$, as otherwise there exists some $i' \in [i_1 + 1, i_2 - 1]$ such that either $w[i_1] \leq w[i'] \leq w[i_2]$ or $w[i_1] \geq w[i'] \geq w[i_2]$, contradicting the assumption that $u$ is a maximum length plateau-rollercoaster in $w$. Repeating this argument for every index $i_j \in \{i_1, i_2, \dots, i_{\vert u \vert}\}$ proves the statement.\looseness=-1
\end{proof}

In order to use $\NEG(w)$, we label each vertex $v_i$ with two lists, $L_{i, \uparrow}, L_{i, \downarrow}$, of length $k$.

\begin{definition}
Let $w\in\Sigma^n$ for $n\in\N$. Define for all $i\in[n]$ and $h\in[k]$, $L_{i, \uparrow}[h]$ as the number of suffixes of plateau-rollercoasters in $w$ of maximum length containing an increasing run where $w[i]$ is the \nth{$h$} unique element in the run. Let $m_{i,j,\uparrow}$ denote the length stored in $R_{w^R}^k[n - i, \uparrow, j] > 0$ if for all $m' \in [m_{i, j, \uparrow}, n]$, we have $R_{w^R}^k[n - i, m', \uparrow, j] = 0$. Define $L_{i, \downarrow}[h]$ analogously.
\end{definition}


With these additional datastructures, we use $\NEG(w)$ to enumerate the set of all maximum length plateau-rollercoasters in a recursive manner. At a high level, the idea is to work backwards, using the tables $L_w^{k, h,\xi}$ to determine which positions in $w$ correspond to the final positions in maximum length plateau-$k$-rollercoasters in $w$. Once the last symbol has been determined, we use the edges in $\NEG(w)$ to determine which preceding positions can be used to construct valid runs ending at the appropriate position. We repeat this until we have determined a valid plateau-rollercoaster, which we then output. By using the the plateau-rollercoaster table, we can check efficiently if a given position is valid:  $R_w^k[i, \xi, h]$ gives us the number partial plateau-$(k, h)_\xi$-rollercoaster ending at $i$, and therefore the number of prefixes of length $L_w^{k, h,\xi}[i]$ ending at $i$. Before, we can prove the main result of this section, we need one lemma relating maximum length plateau-rollercoaster and $P_w$.

\begin{lemma}
    \label{obs:final_position}
    Given $w \in \Sigma^n$, $k\in \N$, and a maximum length plateau-rollercoaster $s = w[i_1] w[i_2] \dots w[i_m]$ in $w$, then $i_m = P_w[n, w[i_m]]$.
\end{lemma}

\ifpaper
\else 
\begin{proof}
    Assume, for the sake of contradiction, that this does not hold. Then, there must be some rollercoaster $s = w[i_1] w[i_2] \dots w[i_m]$ where $i_m \notin P_w[n]$. However, observe that $s w[P_w[n, w[i_m]]]$ is also a plateau-$k$-rollercoaster in $w$ and has longer length than $s$, contradicting the initial assumption, and hence the statement holds.
\end{proof}

\fi

\begin{theorem}
    \label{thm:enumerating_rollercoasters}
    Given a word $w \in \Sigma^n$ for some $n\in\N$ and $k \in \mathbb{N}$, where $m$ is the length of the longest plateau-$k$-rollercoaster in $w$, there exists an algorithm outputting all plateau-$k$-rollercoasters in $w$ of length $m$ with $O(n)$ delay after $O(n \sigma k)$ preprocessing.
\end{theorem}
\begin{proof}
    Before we begin the enumeration, as our preprocessing step, we compute the tables $L_w^{k, h,\xi}$ for every $h \in [1, k]$, $P_w$, the plateau-rollercoaster table $R_w^k$ and the next element graph $\NEG(w) = (V, E)$, requiring $O(n \sigma k)$ time. Further, we use $m$ to denote the length of the longest plateau-$k$-rollercoaster in $w$.\looseness=-1

    To determine the first plateau-rollercoaster, let $i_m$ be some index such that $i_m = P_w[n, x]$, for some $x \in \Sigma$, and $L_w^{k, k,\xi}[i_m] = m$, for some $\xi \in \{\uparrow, \downarrow\}$. 
    By construction, there must exist some plateau-$k$-rollercoaster ending with a $\xi$-run at $i_m$ and further, the preceding position $i_{m - 1}$ must satisfy 
    \begin{itemize}
    \item either $(v_{i_{m - 1}}, v_{i_m}, \xi) \in E$, and $L_w^{k, k,\xi}[i_m] = L_w^{k, h,\xi}[i_{m - 1}] + 1$, for some $h \in \{k - 1, k\}$, 
    \item or $(v_{i_{m - 1}}, v_{i_m}, \rightarrow) \in E$, and $L_w^{k, k,\xi}[i_m] = L_w^{k, k,\xi}[i_{m - 1}] + 1$. 
    \end{itemize}
    We can determine the value of $i_{m - 1}$ by checking each incoming edge $(v_{j}, v_{i_m}, \xi)$ and the edge $(v_j, v_{i_m}, \rightarrow)$ in at most $O(\sigma)$ time. We assume, without loss of generality, that we choose the value of $i_{m - 1}$ satisfying for all $j \in [i_{m - 1} + 1, i_m - 1]$, 
    \begin{itemize}
    \item     either $(v_{i_{m - 1}}, v_{i_m}, \xi) \notin E$, $(v_{i_{m - 1}}, v_{i_m}, \rightarrow) \notin E$, $(v_{i_{m - 1}}, v_{i_m}, \xi) \in E$, and $L_w^{k, h',\xi}[j] + 1 < L_w^{k, k,\xi}[i_m]$, for all  $h' \in \{k - 1, k\}$,
    \item or $(v_{i_{m - 1}}, v_{i_m}, \rightarrow) \in E$ and $L_w^{k, k,\xi}[j] + 1 < L_w^{k, k,\xi}[i_m]$.
    \end{itemize}

    In general, once the value of $i_j$ has been determined, we determine the value of $i_{j - 1}$ as follows. Let us assume that $i_j$ is the \nth{$h$} unique element of a $\xi$-run. If $h \geq k$, then $i_{j - 1}$ must be some index such that 
    \begin{itemize}
    \item  either   $(v_{i_{j - 1}}, v_{i_{j}}, \xi) \in E$ and $L_w^{k, h',\xi}[j - 1] + 1 = L_w^{k, k,\xi}[j]$, for some $h' \in \{k - 1, k\} $, 
    \item or $(v_{i_{j - 1}}, v_{i_{j}}, \rightarrow) \in E$, and $L_w^{k, k,\xi}[j - 1] + 1 = L_w^{k, k,\xi}$. 
    \end{itemize}
    If $h \in [2, k - 1]$, then $i_{j - 1}$ must satisfy 
    \begin{itemize}
    \item   either $(v_{i_{j - 1}}, v_{i_{j}}, \xi) \in E$ and $L_w^{k, h - 1,\xi}[j - 1] + 1 = L_w^{k, h,\xi}[j]$,\\
    \item or $(v_{i_{j - 1}}, v_{i_{j}}, \rightarrow) \in E$ and $L_w^{k, h,\xi}[j - 1] + 1 = L_w^{k, h,\xi}$. 
    \end{itemize}
    Finally, if $h = 1$, then $i_{j - 1}$ must satisfy 
    \begin{itemize}
    \item either $(v_{i_{j - 1}}, v_{i_{j}}, \overline{\xi}) \in E$ and $L_w^{k, h',\xi}[j - 1] + 1 = L_w^{k, k,\xi}[j]$, for some $h' \in \{k - 1, k\}$,
    \item or $(v_{i_{j - 1}}, v_{i_{j}}, \rightarrow) \in E$, and $L_w^{k, h',\xi}[j - 1] + 1 = L_w^{k, 1,\xi}$, for some $h' \in \{k - 1, k\} $.
    \end{itemize}
    In all cases, we can determine this in $O(\sigma)$ time by checking each incoming edge to $v_{i_j}$ in order, terminating once a valid index $i_{j-1}$ has been found. Further, we assume that $i_{j - 1}$ is the largest index satisfying this condition and further that we check each candidate index from largest to smallest.\looseness=-1

    To get the delay in the first output, observe that no index is checked as a candidate to be added to the plateau-rollercoaster more than once. Therefore, as there are at most $n$ possible checks, the total delay is $O(n)$.\looseness=-1

    To determine the next plateau-rollercoaster after outputting the plateau-rollercoaster $s = w[i_1] w[i_2] \dots w[i_m]$, we require an auxiliary set of $m$ counters, $c_1, c_2, \dots, c_m$, where $c_j$ counts the number of plateau-rollercoasters that have been output so far with the suffix $w[i_j] w[i_{j + 1}] \dots w[i_{m}]$. Note that if $c_j < R_w^k[i_j, \xi, h]$, where $w[i_j]$ is the \nth{$h$} (or at least \nth{$h$} if $h = k$) unique symbol in a $\xi$-run in $s$, then there exists some plateau-rollercoaster with the suffix $w[i_j] w[i_{j + 1}] \dots w[i_m]$ that has not been output and otherwise every plateau-rollercoaster with this suffix has been output. Now, let $j$ be the value such that $c_j < R_w^k[i_j, \xi_j, h_j]$ and $c_{j'} = R_w^k[i_{j'}, \xi_{j'}, h_{j'}]$, for every $j' \in [1, j - 1]$ where $i_{f}$ is the \nth{$h_{f}$} unique symbol in a $\xi_f$ run in $s$. We construct the new plateau-rollercoaster by first finding the new value $i_{j - 1}'$ in the same manner as for the first plateau-rollercoaster, with the additional constraint that $i_{j - 1}' < i_{j - 1}$. Once $i_{j - 1}'$ has been determined, we find the value of $i_{j'}$, for every $j' \in [j - 2]$ in the same manner as above, without any additional restriction on the maximum value. Again, by choosing the largest possible value for $i_{j'}'$ each time, we ensure that at most $O(n)$ comparisons are needed, and thus the delay between output is at most $O(n)$. Once the new set of indices has been determined, we output the new plateau-rollercoaster $s = w[i_1'] w[i_2'] \dots w[i_{j - 1}'] w[i_j] w[i_j + 1] \dots w[i_m]$, and update the value of $c_{j'}$ to $1$, if $j' < j$, or to $c_{j'} + 1$ if $j' \geq j$.\looseness=-1
    
    To show correctness, observe that each output corresponds to a unique path in $\NEG(w)$, and thus, by Lemma~\ref{obs:every_path_is_unique}, a unique plateau-rollercoaster. Further, as we output $R_w^k[i_j, \xi, h]$ plateau-rollercoasters with the suffix $w[i_j] w[i_{j + 1}], \dots$, $w[i_{m}]$, when $w[i_j]$ is the \nth{$h$} element on a $\xi$-run, we must output every possible such plateau-rollercoaster, completing the proof.\looseness=-1
\end{proof}

With this proof we conclude the section on searching for the longest plateau-$k$-rollercoaster within a word and move our attention to the problem on finding the longest common plateau-$k$-rollercoaster within a set of words.\looseness=-1

	\section{Longest Common Plateaux Rollercoasters}\label{lcr}
	In this section, we provide tools for finding the longest common plateau-$k$-rollercoaster $s$ in a set of words $\mathcal{W} = \{w_1, w_2, \dots, w_m\}$, i.e., $s$ is a plateau-$k$-rollercoaster in every $w_i$, $i\in[m]$ and there does not exist a longer such plateau-$k$-rollercoaster. As with finding the longest single plateau-rollercoaster, there may be multiple such plateau-rollercoasters. We prove this using the same tools as in Section \ref{sec:results}, namely the lists $L_{w_i}^{k, h,\xi}$, the tables $P_{w_i}$ and rollercoaster tables $R_{w_i}^k$, for every $w_i \in \mathcal{W}, \xi \in \{\uparrow, \downarrow\}$ and $h \in [k]$. Our primary result in this section is an $O(Nk\sigma)$ time algorithm for finding the longest common plateau-$k$-rollercoaster in a given set of words $\mathcal{W}$, integer $k$, defined over an alphabet of size $\sigma$, where $N = \prod_{w \in \mathcal{W}} \vert w \vert$. First, we formally define longest common plateau-$k$-rollercoasters.

\begin{definition}
    Given a finite set $\mathcal{W}\subset\Sigma^{\ast}$, a plateau-$k$-rollercoaster $s$  is \emph{common} to $\mathcal{W}$ if $s$ is a plateau-$k$-rollercoaster of every $w \in \mathcal{W}$. A \emph{longest common plateau-$k$-rollercoaster} of the set $\mathcal{W}$ is a plateau-$k$-rollercoaster $s$ which is common to $\mathcal{W}$ and any plateau-$k$-rollercoaster $s'$ where $\vert s' \vert > \vert s \vert$ is not common to $\mathcal{W}$.\looseness=-1
\end{definition}

For the remainder of this section, let $\mathcal{W}=\{w_1,\ldots,w_n\}\subset\Sigma^{\ast}$ for $n\in\N$ and $k\in\N$.
We solve the longest common plateau-rollercoaster-problem using dynamic programming. The main tool we use is the set of tables $\LCR_{\mathcal{W}}^{k, h,\xi}$.

\begin{definition}
Let $h\in[k]$, and $\xi\in\{\uparrow,\downarrow\}$.
Define the $|w_1|\times\cdots\times|w_n|$ matrix $LCR_{\mathcal{W}}^{k, h,\xi}$ such that $LCR_{\mathcal{W}}^{k, h,\xi}[i_1, \dots, i_m]$ contains the length of the longest common plateau-$(k, h)_{\xi}$-rollercoaster of the words $w_1[1, i_1], w_{2}[1, i_2], \dots,$ $w_{m}[1, i_m]$ that includes $w_1[i_1], w_2[i_2], \dots, w_{m}[i_m]$.\looseness=-1
\end{definition}

The last condition of the definition already contains the conclusion that the longest common plateau-rollercoasters  is empty if it ends in different letters.
We compute the values of $LCR_{\mathcal{W}}^{k, h,\xi}$ similarly to the values of $L_w^{k,h,\xi}$ in Section \ref{sec:results}. For notational brevity, we introduce the table $\mathcal{P}_{\mathcal{W}}$ analogously to $P_w$.
Notice that $\mathcal{P}_{\mathcal{W}}$ may be computed (analogously to Section~\ref{sec:results}) in $O(N \sigma)$ time, where $N = \prod_{w \in \mathcal{W}} \vert w \vert$.

\begin{definition}
	The table $\mathcal{P}_{\mathcal{W}}$ of size $\vert w_1 \vert \times \vert w_2 \vert \times \dots \times \vert w_m \vert \times \sigma$ is defined by $\mathcal{P}_{\mathcal{W}}[i_1, i_2, \dots, i_m, x]$ $=$ $(P_{w_1}[i_1, x]$, $P_{w_2}[i_2, x]$, $\dots$, $P_{w_m}[i_m, x])$.
\end{definition} 

For the above mentioned reason, we consider, without loss of generality, only index tuples $(i_1,\ldots,i_m)$ with 
$w_j[i_j]=w_{j'}[i_{j'}]$ for all $j,j'\in[m]$. The following lemma gives the computation for $\LCR_{\mathcal{W}}^{k,h,\xi}$ for all $h\in[k]$ and $\xi\in\{\uparrow,\downarrow\}$.

\begin{lemma}
    \label{lem:computing_LCR_k_k}
    Given $h\in[k]$ and $(i_1,\ldots,i_m)$ appropriate, we have
    \begin{multline*}
    \LCR_{\mathcal{W}}^{k, k,\uparrow}[i_1, \dots, i_m]=\\
    1 + \max
    \left\{
            \max_{x \in [1, w_1[i_1] - 1]} \{\LCR_{\mathcal{W}}^{k, k - 1,\uparrow}[\mathcal{P}_{\mathcal{W}}[i_1 - 1, , \dots, i_m - 1, x]\},\right.\\
     \left.   \max_{x \in [1, w_1[i_1]]} \{\LCR_{\mathcal{W}}^{k, k,\uparrow}[\mathcal{P}_{\mathcal{W}}[i_1 - 1, , \dots, i_m - 1, x]\}  
    \right\},
    \end{multline*}
    if either $LCR_{\mathcal{W}}^{k, k - 1,\uparrow}[\mathcal{P}_{\mathcal{W}}[i_1 - 1,\dots, i_m - 1, x]) > 0$ for some $x \in [w_1[i_1] - 1]$ or $\max_{x \in [1, w_1[i_1]]} (LCR_{\mathcal{W}}^{k, k,\uparrow}[\mathcal{P}_{\mathcal{W}}[i_1 - 1, \dots, i_m - 1, x]) > 0$ for some $x \in [w_1[i_1]]$,
    \begin{multline*}
    \LCR_{\mathcal{W}}^{k, h,\uparrow}[i_1, \dots, i_m]= \\
     1 + \max \left\{ 
                    \max_{x \in [1, w_1[i_1] - 1]} \{\LCR_{\mathcal{W}}^{k, h - 1,\uparrow}[\mathcal{P}_{\mathcal{W}}[i_1 - 1, \dots, i_m - 1, x]\},\right.\\
          \left.  \LCR_{\mathcal{W}}^{k, h,\uparrow}[\mathcal{P}_{\mathcal{W}}[i_1 - 1, \dots, i_m - 1, w_1[i_1]]
        \right\}
    \end{multline*}
    if either $LCR_{\mathcal{W}}^{k, h - 1,\uparrow}[\mathcal{P}_{\mathcal{W}}[i_1 - 1, i_m - 1, x]] > 0$ for some $x \in [w_1[i_1] - 1]$, or $LCR_{\mathcal{W}}^{k, h,\uparrow}[\mathcal{P}_{\mathcal{W}}[i_1 - 1, dots, i_m - 1, w_1[i_1]]> 0$ for some $x \in [w_1[i_1]]$, and $0$ otherwise, and finally 
    \[
    \LCR_{\mathcal{W}}^{k, 1,\uparrow}[i_1, \dots, i_m]=\LCR_{\mathcal{W}}^{k, k,\downarrow}[i_1, \dots, i_m],
    \]
    if the respective plateau-$(k,1)_\xi$-rollercoasters are not unary.
    
    \medskip
    
    The values for $\xi=\downarrow$ can be computed analogously.
\end{lemma}

\ifpaper
\else 
\begin{proof}
We only prove the first claim since all other claims follow analogously.
    Note that if the condition is not met, then there is no plateau-$(k, k - 1)_{\uparrow}$-rollercoaster that is common to $\{w_1[1, i_1 - 1], \dots, w_m[1, i_m - 1]\}$, and thus there does not exist any plateau-$(k, k)_{\uparrow}$-rollercoaster that is common to the set $\{w_1[1, i_1], \dots, w_m[1, i_m]\}$.
    
    Now, observe that the longest plateau-$(k, k)_{\uparrow}$-rollercoaster ending at $w_1[i_1]$, $\dots$, $w_{m}[i_m]$ must be preceded by either a plateau-$(k, k - 1)_{\uparrow}$-rollercoaster ending with some letter strictly smaller than $w_1[i_1]$ (equivalently, $w_j[i_j]$ for any $j \in [m]$). or a plateau-$(k, k)_{\uparrow}$-rollercoaster ending with some letter smaller than or equal to $w_1[i_1]$. Further, for any letter $x \in \Sigma$, the longest common plateau-$(k, h)_{\uparrow}$-rollercoaster in $\{w_1[1, i_1], \dots, w_m[1, i_m]\}$ ending with the letter $x$ must have length $\LCR_{\mathcal{W}}^{k, h,\uparrow}[\mathcal{P}_{\mathcal{W}}[i_1 - 1, \dots, i_m - 1, x]]$, as given any plateau-$(k, h)_{\uparrow}$-rollercoaster that is common to $\{w_1[1, i_1]$, $\dots$, $w_m[1, i_m]\}$ ending with $x$ at some position before $P_{w_j}[i_j - 1, x]$ one of equivalent length can be formed by swapping the last position used in $w_j$ with $P_{w_j}[i_j - 1, x]$. Thus, this statement holds.
\end{proof}

\fi

Now, we present the main result of this section, the computation of the longest common plateau-$k$-rollercoaster
of a given finite set $\mathcal{W}\subset\Sigma^{\ast}$.

%
%

\begin{theorem}
    \label{thm:LCR}
    The set of tables $\LCR_{\mathcal{W}}^{k, h,\xi}$ and thus the length of the longest common plateau-$k$-rollercoaster, can be computed, for every $h \in [k], \xi \in \{\uparrow, \downarrow\}$, in $O(N k \sigma)$ time, where $N = \prod_{w \in \mathcal{W}} \vert w \vert$. (The pseudocode for the algorithm can be found in Appendix \ref{app:pseudocode}.)
\end{theorem}

\begin{proof}
    Lemma \ref{lem:computing_LCR_k_k} provide the outlines for a recursive approach to computing the tables $LCR_{\mathcal{W}}^{k, h,\xi}$. As a base case, the values of $LCR_{\mathcal{W}}^{k, h,\xi}[i_1, \dots, i_m]$ can be set to $0$ for any set $i_1, \dots, i_m$ where $w_j[i_j] \neq w_{\ell}[i_{\ell}]$ for some pair $j, \ell \in [m]$, requiring $O(N k)$ time. Similarly, we can determine if the value of $LCR_{\mathcal{W}}^{k, 1,\xi}[1, \dots, 1]$ is $1$, if $w_1[1] =  \dots = w_m[1]$, or $0$ otherwise, for both values of $\xi \in \{\uparrow, \downarrow\}$. Finally, we set the value of $LCR_{\mathcal{W}}^{k, h,\xi}[1, \dots, 1]$ to $0$ for every $h \in [2, k]$.

    In the general case, in order to compute the value of $LCR^{k, h,\xi}_{\mathcal{W}}[i_1, \dots, i_m]$, let us assume that the values of $\LCR^{k, h',\xi}_{\mathcal{W}}[i_1', \dots, i_m']$ has been computed for every $i_1' \in [i_1],  \dots, i_m' \in [i_m]$, $h' \in [k]$, and $\xi' \in \{\uparrow, \downarrow\}$, other than $(i_1', \dots, i_m') = (i_1, \dots, i_m)$. Further, we assume that, if $h = 1$, the value $\LCR^{k, k,\xi}_{\mathcal{W}}[i_1, \dots, i_m]$ has already been computed, noting that the value $\LCR^{k, k,\xi}_{\mathcal{W}}[i_1, \dots, i_m]$ does not depend upon $\LCR^{k, 1,\xi}_{\mathcal{W}}[i_1, \dots, i_m]$. From Lemma \ref{lem:computing_LCR_k_k},  $\LCR^{k, h,\xi}_{\mathcal{W}}[i_1, \dots, i_m]$ can be computed in $O(\sigma)$ time using one of the the given formulae.

    As there are $N$ entries in $\LCR^{k, h,\xi}_{\mathcal{W}}$, $2 k$ tables $\LCR^{k, 1,\xi}_{\mathcal{W}}$, $\dots$, $\LCR^{k, k,\xi}_{\mathcal{W}}$, and the complexity of computing each entry is $O(\sigma)$ time, the total complexity of computing every table is $O(N k \sigma)$.
    
    The length can be obtained immediately from $\LCR_{\mathcal{W}}^{k,h,\xi}$.
\end{proof}

	\section{Conclusion}\label{conc}
	Within this work we introduced and investigated the notion of plateau-$k$-roller\-coaster as a natural extension to $k$-rollercoaster relaxing the strictly in/decreasing runs to weakly in/decreasing runs. First, we gave an $O(n\sigma k)$-algorithm to determine the longest plateau-$k$-rollercoaster for an $n$-length word over an $\sigma$-letter alphabet. Extending this idea, we introduced a \emph{rollercoaster table} which allows for enumerating all longest plateau-$k$-rollercoaster of a word of length $n$ with $O(n)$ delay after $O(n\sigma k)$ preprocessing.
Second, we presented an algorithm to search for the longest common plateau-$k$-rollercoaster within a set of words. Via a dynamic programming approach, the longest common plateau-$k$-rollercoaster can be computed in $O(Nk\sigma)$ time where $N$ is the product of all word lengths within the set.
For further research, one might proceed with algorithmical studies on, e.g., the shortest common supersequence of a set of words that is a plateau-$k$-rollercoaster. 

	%
	%
	%
	 \bibliographystyle{splncs04}
	 \bibliography{mybibliography}
\newpage
  \appendix

  \section{Pseudo Code}
  \label{app:pseudocode}
    For all algorithms, we assume that every positions starts at $0$.
\begin{algorithm}
\textbf{Input:} $w \in \Sigma^n$, $k \in \mathbb{N}$\\
\textbf{Output:} rollercoaster table $R$ for $w$ and $k$
\begin{lstlisting}[mathescape=true,basicstyle=\small]
$R^k_w$ = empty table of size $n \times 1 \times k$
Initialise $L_{w}^{k, h,\xi}$, $\forall \xi \in \{\uparrow, \downarrow\}, h \in [1, k]$
Compute $P_w$
// Initialisation of for in/decreasing starting runs
for $i\in[n]$, $\xi\in\{\uparrow,\downarrow\}$
  $R_w^k[i, \xi, 1]= 1$ 
// Partial RC, starting by $2$, (no partial RC starts in $1$); each $R_w^k[i, \xi, h]$ is computed for every $h \in [1, k]$ and $\xi \in \{\uparrow, \downarrow\}$ before increasing $i$. Automatically compute $R_w^k[i, \xi, 1]$ by setting it to $R_w^k[i, \overline{\xi}, k]$ once that value has been computed.
for $i \in [2, n], h \in [2, k]$:
  $\PP^{k, h,\uparrow}_{w}[i]= \{P_w[x, i - 1] \mid x \in [1, w[i] - 1], L_{w}^{k, h - 1,\uparrow}[P_w[i - 1, x]]= L_{w}^{k, h,\uparrow}[i] - 1\}$
  $\PP^{k, h,\downarrow}_{w}[i]= \{P_w[x, i - 1] \mid x \in [w[i] + 1, \sigma], L_{w}^{k, h - 1,\downarrow}[P_w[i - 1, x]]= L_{w}^{k, h,\downarrow}[i] - 1\}$
  for $i' \in \PP^{k, h,\xi}_{w}[i]$, $\xi\in\{\uparrow,\downarrow\}$
    $R^k_w[i, \xi, h] = R^k_w[i, \xi, h] + R^k_w[i', \xi, h - 1]$
  if $h \neq k$
     if $L_{w}^{k, h,\uparrow}[i] - 1 = L_{w}^{k, h,\uparrow}[P_w[i - 1, w[i]]]$:
        $R^k_w[i, \uparrow, h] = R^k_w[i, \uparrow, h] + R^k_w[P_w[i - 1, w[i]], \uparrow, h]$
     if $L_{w}^{k, h,\downarrow}[i] - 1 = L_{w}^{k, h,\downarrow}[P_w[i - 1, w[i]]]$:
        $R^k_w[i, \downarrow, h] = R^k_w[i, \downarrow, h] + R^k_w[P_w[i - 1, w[i]], \downarrow, h]$
  else
     $\PP^{k, h,\uparrow}_{w}[i]'= \{P_w[x, i - 1] \mid x \in [1, w[i]], L_{w}^{k, h,\uparrow}[P_w[i - 1, x]]= L_{w}^{k, h,\uparrow}[i] - 1\}$
     $\PP^{k, h,\downarrow}_{w}[i]= \{P_w[x, i - 1] \mid x \in [w[i], \sigma], L_{w}^{k, h,\downarrow}[P_w[i - 1, x]]= L_{w}^{k, h,\downarrow}[i] - 1\}$
     for $i' \in \PP^{k, h,\uparrow}_{w}[i]$:
       $R^k_w[i, \uparrow, h] = R^k_w[i, \uparrow, h] + R^k_w[i', \uparrow, h]$
     for $i' \in \PP^{k, h,\downarrow}_{w}[i]$:
       $R^k_w[i, \downarrow, h] = R^k_w[i, \downarrow, h] + R^k_w[i', \downarrow, h]$
       $R^k_w[i, \uparrow, 1] = R^k_w[i, \downarrow, k]$
       $R^k_w[i, \downarrow, 1] = R^k_w[i, \uparrow, k]$
return $R^k_w$          
\end{lstlisting} 

\caption{ComputeR}\label{alg:computing_R_w}
\end{algorithm}


\begin{algorithm}
    \textbf{Input:} Set of words $\mathcal{W} \in (\Sigma^*)^m$, $k \in \mathbb{N}$\\
    \textbf{Output:} The set $LCR_{\mathcal{W}}^{k, h,\xi}$ for the set $\mathcal{W}$ and $k$, for every $\xi \in \{\uparrow, \downarrow\}$ and $h \in [1, k]$.
    \begin{lstlisting}[mathescape=true]
Compute $\mathcal{P}_{\mathcal{W}}$
Initialise $LCR_{\mathcal{W}}^{k, h,\xi}$ with $0$ for every position, for every $h \in [1, k], \xi \in \{\uparrow, \downarrow\}$.
$LCR_{\mathcal{W}}^{k, 1,\uparrow}[1, 1, \dots, 1] = 1$
$LCR^{k, 1,\downarrow}_{\mathcal{W}}[1, 1, \dots, 1] = 1$
for $i_1 \in [1, \vert w_1 \vert], i_2 \in [1, \vert w_2 \vert], \dots, i_m \in [1, \vert w_m \vert]$
    if $w_1[i_1] = w_2[i_2] = \dots w_m[i_m]$
        for $h \in [2, k - 1]$
            $a_{\uparrow} = \max_{x \in [1, w_1[i_1]]- 1} LCR_{\mathcal{W}}^{k, h - 1,\uparrow}[\mathcal{P}_{\mathcal{W}}[i_1 - 1, i_2 - 1, \dots, i_m - 1]]$
            $a_{\downarrow} = \max_{x \in [w_1[i_1]] + 1, \sigma} LCR_{\mathcal{W}}^{k, h - 1,\downarrow}[\mathcal{P}_{\mathcal{W}}[i_1 - 1, i_2 - 1, \dots, i_m - 1]]$
            $A_{\uparrow} = \max (a_{\uparrow}, LCR_{\mathcal{W}}^{k,h ,\uparrow}[\mathcal{P}_{\mathcal{W}}[i_1 - 1, i_2 - 1, \dots, i_m - 1, w_1[i_1]]])$
            $A_{\downarrow} = \max (a_{\downarrow}, LCR_{\mathcal{W}}^{k, h,\downarrow}[\mathcal{P}_{\mathcal{W}}[i_1 - 1, i_2 - 1, \dots, i_m - 1, w_1[i_1]]])$
            if $A_{\uparrow} > 0$
                $LCR^{k, h,\uparrow}_{\mathcal{W}}(i_1, i_2, \dots, i_m) = A_{\uparrow} + 1$
            if $A_{\downarrow} > 0$
                $LCR^{k, h,\downarrow}_{\mathcal{W}}(i_1, i_2, \dots, i_m) = A_{\downarrow} + 1$
        $a_{\uparrow}' = \max_{x \in [1, w_1[i_1]]- 1} LCR_{\mathcal{W}}^{k, k - 1,\uparrow}[\mathcal{P}_{\mathcal{W}}[i_1 - 1, i_2 - 1, \dots, i_m - 1]]$
        $a_{\downarrow}' = \max_{x \in [w_1[i_1]] + 1, \sigma} LCR_{\mathcal{W}}^{k, k - 1,\downarrow}[\mathcal{P}_{\mathcal{W}}[i_1 - 1, i_2 - 1, \dots, i_m - 1]]$
        $b_{\uparrow} = \max_{x \in [1, w_1[i_1]]} LCR^{k, k,\uparrow}_{\mathcal{W}}[\mathcal{P}_{\mathcal{W}}[i_1 - 1, i_2 - 1, \dots, i_m - 1, x]]$
        $b_{\downarrow} = \max_{x \in [w_1[i_1], \sigma]} LCR^{k, k,\downarrow}_{\mathcal{W}}[\mathcal{P}_{\mathcal{W}}[i_1 - 1, i_2 - 1, \dots, i_m - 1, x]]$
        $B_{\uparrow} = \max(a_{\uparrow}', b_{\uparrow})$
        $B_{\downarrow} = \max(a_{\downarrow}', b_{\downarrow})$
        if $B_{\uparrow} > 0$
            $LCR^{k, k,\uparrow}_{\mathcal{W}}(i_1, i_2, \dots, i_m) = B_{\uparrow} + 1$
        if $B_{\downarrow} > 0$
            $LCR^{k, k,\downarrow}_{\mathcal{W}}(i_1, i_2, \dots, i_m) = B_{\downarrow} + 1$
Return $LCR^{k, 1,\uparrow}_{\mathcal{W}}, LCR^{k, 2,\uparrow}_{\mathcal{W}}, \dots, LCR^{k, k,\uparrow}_{\mathcal{W}}, LCR^{k, 1,\downarrow}_{\mathcal{W}}, LCR^{k, 2,\downarrow}_{\mathcal{W}}, \dots, LCR^{k, k,\downarrow}_{\mathcal{W}}$
    \end{lstlisting}
    \caption{Construct$LCR$}\label{alg:constructing_LCR}
\end{algorithm}

    \newpage
  \section{Extended Example}
  \label{app:example}
  Within this section, we will exemplary generate the tables $P_w$, $L_w^{k,h,\xi}$ for the word $w = 871264435161$ of length $|w| = n = 12$.
Its longest plateau-$3$-rollercoasters are given by $8712644311$ and $8712644356$. Furthermore, note that $L_w^{k,1,\xi}$ and $L_{w}^{k,k,\bar{\xi}}$ do either represent unary plateau-rollercoasters or contain equal values (cf. Remark~\ref{rem:kkandk1rollercoasters}).
\begin{table}[h!]
	\centering
	\begin{tabular}{c || c c c c c c c c} 
		$n\backslash \sigma$ & 1 & 2 & 3 & 4 & 5 & 6 & 7 & 8 \\ [0.5ex] 
		\hline\hline
		1 & - & - & - & - & - & - & - & 1\\ 
		2 & - & - & - & - & - & - & 2 & 1\\
		3 & 3 & - & - & - & - & - & 2 & 1\\
		4 & 3 & 4 & - & - & - & - & 2 & 1\\
		5 & 3 & 4 & - & - & - & 5 & 2 & 1\\
		6 & 3 & 4 & - & 6 & - & 5 & 2 & 1\\
		7 & 3 & 4 & - & 7 & - & 5 & 2 & 1\\
		8 & 3 & 4 & 8 & 7 & - & 5 & 2 & 1\\
		9 & 3 & 4 & 8 & 7 & 9 & 5 & 2 & 1\\
		10 & 10 & 4 & 8 & 7 & 9 & 5 & 2 & 1\\
		11 & 10 & 4 & 8 & 7 & 9 & 11 & 2 & 1\\
		12 & 12 & 4 & 8 & 7 & 9 & 11 & 2 & 1\\
		[1ex] 
		\hline
	\end{tabular}
		\quad
		\begin{tabular}{c | c c c c c c c c c c c c} 
			$i \in [n]$ & 1 & 2 & 3 & 4 & 5 & 6 & 7 & 8 & 9 & 10 & 11 & 12\\ [0.5ex] 
			\hline\hline
			$L_w^{3,1,\downarrow}$ & 1 & 1 & 1 & 1 & 5  & 5 & 6 & 5 & 7 & 1 & 10 & 1\\ 
			$L_w^{3,1,\uparrow}$ & 1 & 1 & 3 & 3 & 3  & 4 & 5 & 8 & 4 & 9 & 4 & 10\\ 
			$L_w^{3,2,\downarrow}$ & 0 & 2 & 2 & 2 & 2  & 6 & 7 & 7 & 4 & 8 & 3 & 9\\ 
			$L_w^{3,2,\uparrow}$ & 0 & 0 & 0 & 4 & 4 & 4 & 5 & 5 & 9 & 0 & 9 & 0 \\
			$L_w^{3,3,\downarrow}$ & 0 & 0 & 3 & 3 & 3  & 4 & 5 & 8 & 4 & 9 & 4 & \textbf{10}\\ 
			$L_w^{3,3,\uparrow}$ & 0 & 0 & 0 & 0 & 5 & 5 & 6 & 5 & 7 & 0 & \textbf{10} & 0\\
			[1ex] 
			\hline
		\end{tabular}
	\caption{The values of $P_w$ and $L_w^{k,h,\xi}$ (the bold values mark the longest rollercoaster within $w$).}
	\label{table:P}
\end{table}

Further, we give some values of the rollercoaster table of $w$:
\begin{itemize}
	\item $R^3_w[(11,\uparrow,3)] = R^3_w[(12,\downarrow,3)] = 1$, and
	\item $R^3_w[(3,\downarrow,2)] = 2$ since both $81$ and $71$ are plateau-$(3,2)_\downarrow$-rollercoasters that end in $w[3] = 1$.
\end{itemize}

\end{document}